\documentclass[sigplan,10pt]{acmart}
\makeatletter
\def\@ACM@checkaffil{
    \if@ACM@instpresent\else
    \ClassWarningNoLine{\@classname}{No institution present for an affiliation}%
    \fi
    \if@ACM@citypresent\else
    \ClassWarningNoLine{\@classname}{No city present for an affiliation}%
    \fi
    \if@ACM@countrypresent\else
        \ClassWarningNoLine{\@classname}{No country present for an affiliation}%
    \fi
}
\makeatother

\acmConference[]{}{}{}
\acmYear{}
\acmISBN{} 
\acmDOI{} 
\startPage{1}
\setcopyright{none}
\usepackage{amsmath,amsfonts}
\usepackage{graphicx}
\usepackage{textcomp}
\usepackage{xcolor}
\def\BibTeX{{\rm B\kern-.05em{\sc i\kern-.025em b}\kern-.08em
    T\kern-.1667em\lower.7ex\hbox{E}\kern-.125emX}}

\usepackage{etoolbox}
\makeatletter
\patchcmd{\maketitle}
{\@maketitle}
{\@maketitle\vspace{-6em}}
{}
{}
\makeatother
\usepackage{tikz}
\usepackage{filecontents}
\usepackage{color, colortbl}
\usepackage{algorithmicx}
\usepackage{algorithm}
\usepackage[noend]{algpseudocode}
\usepackage{siunitx}
\usepackage{balance}
\usepackage{makecell}
\usepackage[skip=4pt]{caption}
\usepackage{diagbox}
\usepackage{float}
\usepackage{hyperref}
\usepackage{siunitx}
\usepackage{threeparttable}
\usepackage{url}
\usepackage{verbatim} 
\usepackage{xspace}
\usepackage{totpages}
\usepackage{varwidth}
\usepackage{subfigure}
\usepackage{makecell}
\usepackage{multirow}
\usepackage{flushend}                    
\usepackage[shortlabels]{enumitem}
\usepackage[noend]{algpseudocode}
\usepackage{cleveref}

\newcommand{\sysname}[0]{\texttt{Xenos}\xspace}
\newcommand{\dsysname}[0]{\texttt{d-Xenos}\xspace}

\newcommand{\hwsay}[1]{} 

\newcommand{\para}[1]{\noindent \textbf{#1 }}

\begin{document}

\title[\sysname{}: Dataflow-Centric Optimization to  Accelerate Model Inference on Edge Devices]{\sysname{}: Dataflow-Centric Optimization to  Accelerate Model Inference on Edge Devices}

\author{Runhua Zhang$^*$,
Hongxu Jiang$^*$,
Fangzheng Tian$^*$,
Jinkun Geng$^+$,
Xiaobin Li$^*$,
Yuhang Ma$^*$,
Chenhui Zhu$^*$,
Dong Dong$^*$,
Xin Li$^*$,
Haojie Wang$^\dagger$
}

\affiliation{%
  \institution{$^*$Beihang University, $^+$Stanford University,$^\dagger$Tsinghua University}
}

\sloppypar
\begin{abstract}

Edge computing has been emerging as a popular scenario for model inference. However, the inference performance on edge devices (e.g., Multi-Core DSP, FGPA, etc.) suffers from inefficiency due to the lack of highly optimized inference frameworks. Previous model inference frameworks are mainly developed in an operator-centric way, which provides insufficient acceleration to edge-based inference. Besides, the operator-centric framework incurs significant costs for continuous development and maintenance. 

In this paper, we propose \sysname, which can automatically conduct dataflow-centric optimization of the computation graph and accelerate inference in two dimensions. Vertically,  \sysname develops operator linking technique to improve data locality by restructuring the inter-operator dataflow. Horizontally, \sysname develops DSP-aware operator split technique to enable higher parallelism across multiple DSP units.
Our evaluation proves the effectiveness of vertical and horizontal dataflow optimization, which reduce the inference time by 21.2\%--84.9\% and 17.9\%--96.2\% , respectively. 
Besides, \sysname also outperforms the widely-used TVM by 3.22$\times$--17.92$\times$. Moreover, we extend \sysname to a distributed solution, which we call \dsysname.  \dsysname employs multiple edge devices to jointly conduct the inference task and achieves a speedup of 3.68$\times$--3.78$\times$ compared with the single device.

\end{abstract}

\ccsdesc[500]{Software and its engineering~General programming languages}
\ccsdesc[300]{Social and professional topics~History of programming languages}

\keywords{edge devices inference, dataflow-centric optimization, DNN optimization,  data locality}  

\maketitle

\section{Introduction}


Edge devices are widely applied nowadays and becoming a prevalent scenario for deep learning applications~\cite{icnp17-edge-computing,sec18-edge-computing,infocom-20-edge-computing}. These edge devices have heterogeneous configurations of hardware resources (e.g. memory, computation, etc) and usually require real-time responsiveness, i.e. the duration of the inference cannot take too long. Existing solutions (e.g. TVM~\cite{osdi18-tvm}) execute model inference inefficiently on these platforms, and fail to satisfy the responsiveness requirement.


With a deep dive into numerous typical model inference workflows, we have identified two main reasons for the inference inefficiency.

(1) \textbf{Inefficient dataflow scheduling.} The dataflow scheduling of model inference can seriously spoil memory locality. Take the typical CNN inference as an example, after completing the inference of each layer, the computation operators output the feature maps to the shared memory region. These feature map elements will serve as the input for the next layer, and be fed to multiple DSP units for the following inference computation. However, there is a mismatch between the data layout (output from the prior layer) and the data access sequence (required by the next layer). In other words, DSP units are not reading in the sequential order as what was written previously. Therefore, while reading the feature maps, DSP units suffer from bad data locality and require \emph{unnecessarily} much read operation, which leads to non-trivial overheads and prolongs the inference time. 

(2) \textbf{Hardware-Oblivious parallelism.} Different edge devices are usually equipped with heterogeneous computation resources and memory hierarchies. A fixed model partition scheme simply ignores the resource conditions and fails to fit the memory hierarchy and/or cannot fully utilize the computation resource. During the model inference process, parameters are frequently swapped in and out. Only a few digital signal processing (DSP) computing units (DSP cores/slices~\footnote{ Mutli-Core DSP uses the term ``DSP core'' whereas FPGA uses ``DSP slice''. We use ``DSP unit'' as the general term in the following description.}) are active and undertaking the computation tasks, whereas the majority remains idle, waiting for the dependent data. Such partition schemes can waste much computation power and yield no satisfying performance. 

We have studied the existing frameworks (e.g.,  TVM~\cite{osdi18-tvm}, TASO~\cite{jia2019taso} and PET~\cite{osdi21_pet}), but find that they provide very limited optimization regarding the aforementioned two issues. Such optimization proves to be insufficient in edge-based inference. First, the typical frameworks use simple enumeration-based search strategy to try different combinations of operator fusion/split of operators and finally decide a scheme to optimize the computation graph. The search strategy is oblivious to the resource conditions of the edge device and becomes inefficient as the number of operators grows. For example, the enumeration-based search strategy in TASO and Pet can only work with less than 5 operators in practice, which restricts their application scope. Second, the existing optimization frameworks ignores the data locality during the inference process, leading to costly cache misses.




Motivated by the drawbacks of existing optimization frameworks, we develop a novel solution, \sysname, to accelerate the model inference on edge devices. Contrary to the existing frameworks, \sysname is built based on dataflow-centric optimization instead of operator-centric optimization. Specifically, \sysname is able to save the overheads in two aspects ignored by prior works. 

First (Vertically), \sysname restructures the inference dataflow between adjacent operators and develops operator linking technique to optimize the computation graph. Before running the inference model, \sysname scans the whole computation graph and modify the dataflow between adjacent operators. The customized dataflow information is written into the metadata of the computation graph and fed into \sysname' inference engine. During runtime, the inference engine can write the intermediate result (output from the previous operator) in the same order as read by the subsequent operator. Thus, the inference process yields much better data locality.    

Second (Horizontally), \sysname takes hardware information (DSP units and memory hierarchy) into account and develop DSP-aware operator split technique to manage the dataflow across DSP units. Instead of enumerating every possible scheme of operator split (like TASO and PET), \sysname heuristically partitions the feature map across DSP units for high parallelism, and splits the operator parameters to fit into the private L2 memory for efficient data access. Thus, \sysname can decide a high-performance scheme to deploy the model to the edge device much faster.

We summarize our contributions as below.

\begin{itemize}[leftmargin=*, nosep]
    \item \textbf{Framework}. \sysname is a complete end-to-end framework which focuses on dataflow-centric optimization instead of operator-centric optimization. From the perspective of performance, our approach can conduct more in-depth optimization for edge-based model inference (i.e.,vertical and horizontal dataflow optimization), which are ignored by prior works,  From the perspective of maintenance and development, our approach optimizes computation graphs without introducing new operators, which saves much programming effort in the continuous development.
    
    
    
        

    \item \textbf{Distributed inference}. We also extend \sysname from single-node inference to distributed inference, which we call \dsysname. \dsysname targets at large-volume inference workload which cannot be handled by single edge device efficiently. \dsysname incorporates the bandwidth-optimal ring all-reduce algorithm for parameter synchronization and conduct inference in a model-parallel way, which effectively accelerates the inference computation. 
    \item \textbf{Evaluation}. We conduct comparative experiments on different platforms showing that \sysname{} can reduce the inference time by 21.2\%--84.9\% and 17.9\%--96.2\%, respectively. \sysname{} also outperforms TVM by 3.22$\times$--17.92$\times$. Regarding distributed inference, \dsysname achieves the a speedup by 3.68$\times$--3.78$\times$ compared with the single-device baselines. 
    
\end{itemize}

\vspace{-0.5cm}
\section {Background and Motivation}
\label{sec-back}

\subsection{Model Inference Workflow}
\begin{figure}[!t]
	\centering
	\includegraphics[width=0.48\textwidth]{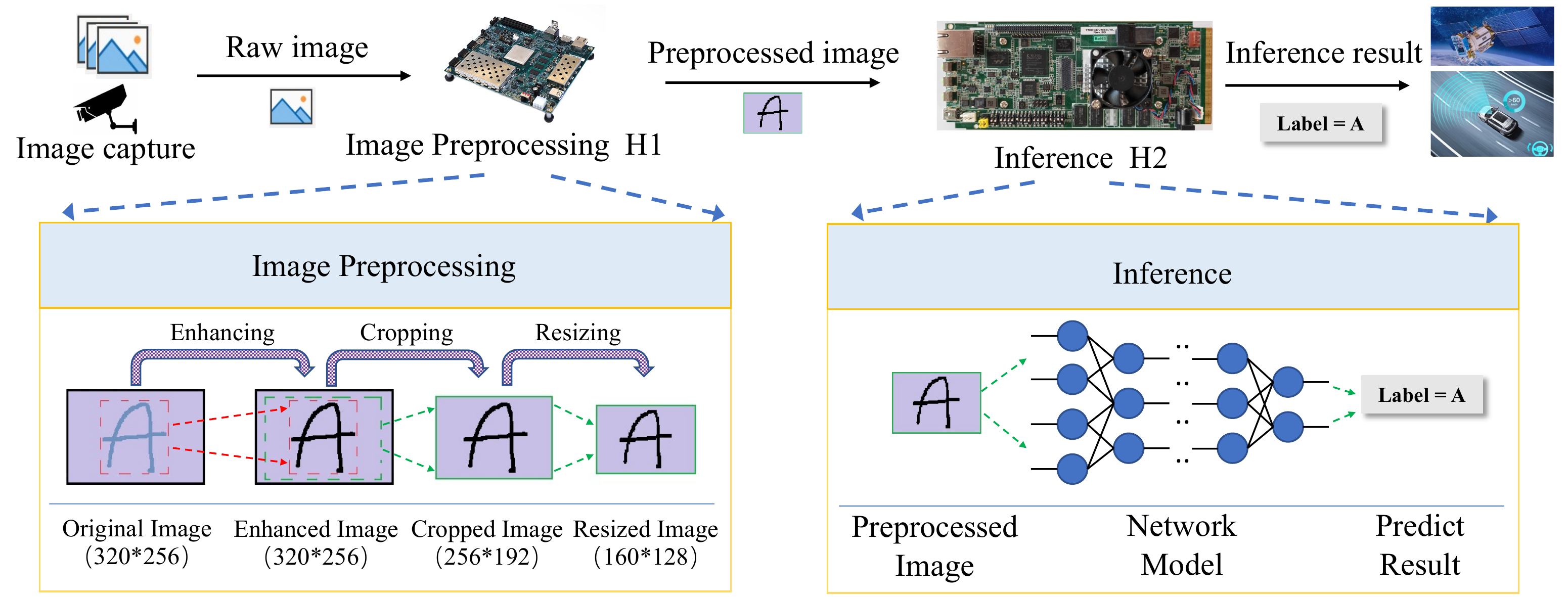}
	\caption{The full stack of an inference workflow.} 
	\vspace{-0.5cm}
	\label{inference}
\end{figure}

Figure~\ref{inference} illustrates the typical workflow for model inference on edge devices, which includes three main parts, i.e., the image acquisition module, image preprocessing module, and inference module.
The image acquisition module collects images either from image capture devices, or from user inputs. 
Then these images will be preprocessed on hardware H1 to fit requirements of the inference module, including size adjustment and image enhancement. 
Finally these preprocessed images will be sent to inference module on hardware H2, which will output the final inference result for applications.

The workflow should be executed very fast to satisfy the requirement of real-time responsiveness.
Among the three modules, the inference module tend to be the latency bottleneck, which typically takes over 60\% of the overall execution time according to our experiments.
The inference module thus becomes the key to improve the system responsiveness, where \sysname is applied for acceleration.

\subsection{Data Locality in Inference Computation}
\label{sec-dataflow-inefficiency}
While executing the inference computation, the DSP units usually read the data in a non-sequential order due to the data layout mismatch for different operators. We use an example of a depthwise separable convolution, which consists of a depthwise convolution followed by a pointwise convolution, to illustrate this.

\begin{figure}[h]
	\centering
	\includegraphics[width=2.5in]{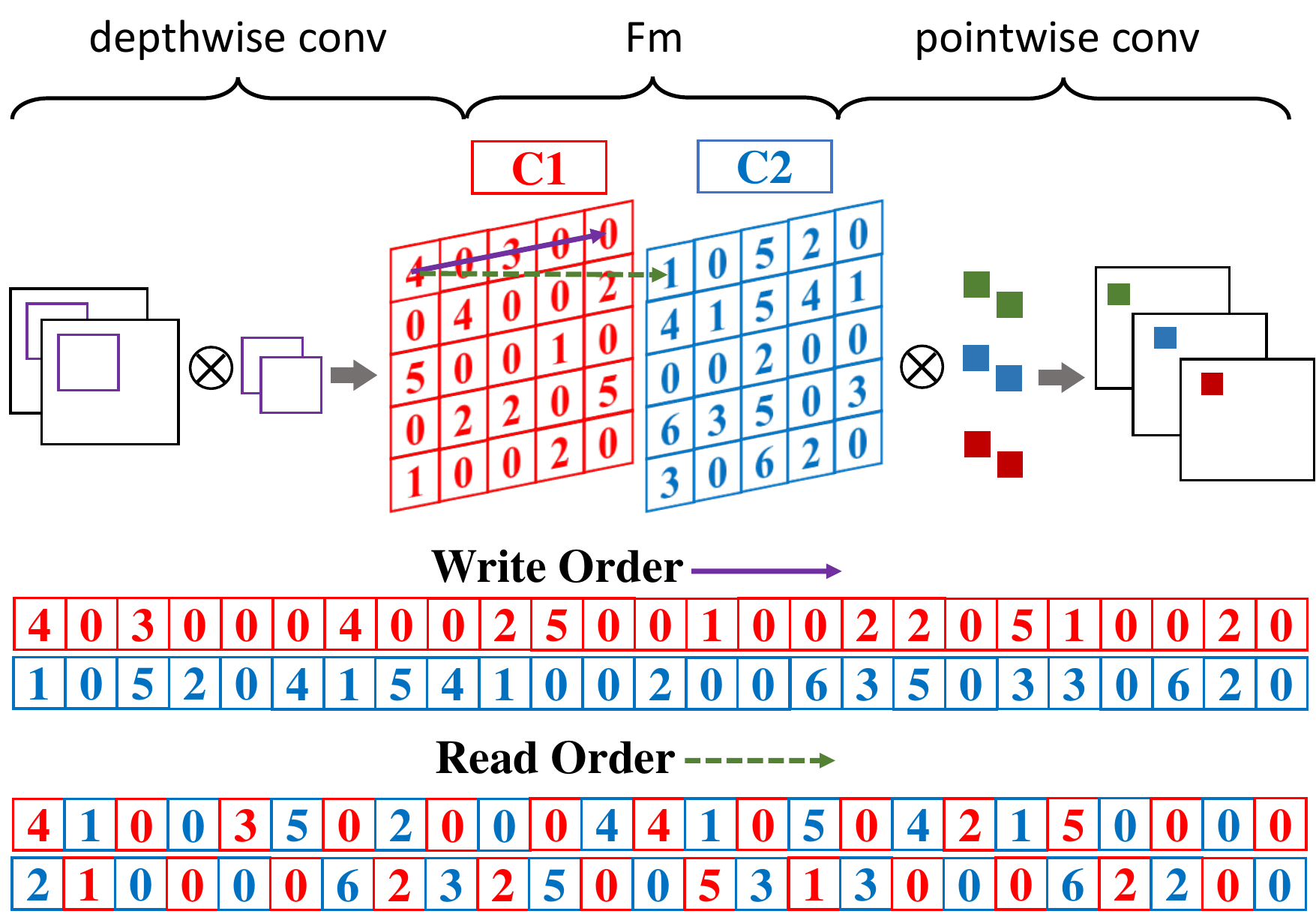}
	\caption{Inefficient dataflow scheduling with bad locality}
	\label{Inefficient-dataflow} 
	\vspace{-0.5cm}
\end{figure}
In \Cref{Inefficient-dataflow}, feature map $Fm$, which has 2 channels $C1$ and $C2$, is the output of a depthwise convolution and the input of a pointwise convolution.
The depthwise convolution write $Fm$ in a width-first order, while the pointwise convolution read $Fm$ in a channel-first order, leading to data layout mismatch between these two operators.
Thus, simply partitioning the model will cause the inference process to suffer from poor data locality, because each DSP unit requires different data blocks when writing and reading the feature maps.

\subsection{Hardware Resource Heterogeneity}
\label{sec-mismatch-memory}


When executing a model inference task on edge devices, it will be efficient to place the whole model in high-level memory (e.g., L2 memory).
Such strategy is non-trivial in practice due to the limited memory on edge devices.
Although existing works~\cite{howard2017mobilenets,he2018amc,polino2018model} have proposed different pruning approaches, the model size is still too large to fit into the memory hierarchy on edge devices, leading to serious performance degrade for inference tasks.
For example, as a lightweight inference model designed for edge hardware, MobileNet~\cite{howard2017mobilenets} still has many layers whose sizes of feature maps and kernels are larger than the size of L2 memory (e.g., 512KB on TMS320C6678) or the size of shared memory (e.g., 4MB on TMS320C6678).
Simply executing this model without partition will lead to significant inefficiency.


As edge devices can be various and possess heterogeneous configurations of resources running models with different sizes, there is no "one-size-fits-all" partition scheme to fit all scenarios. 
Manually-tuned scheme highly relies on human expertise, which can be very inefficient and causes a long deployment period. 
Existing automated solutions, such as TASO~\cite{sosp19_taso} and PET~\cite{osdi21_pet}, enumerate every possible partition scheme in a large search space, which prolongs the deployment time.
Moreover, these frameworks are limited by graph size due to the complexity of their algorithm, which make them miss many optimization opportunities.
Therefore, we are motivated to design vertical dataflow optimization, which takes the hardware resource information into account, and automatically generates a desirable partition scheme to both fully utilize the computation power and well fit into the memory hierarchy. 

\subsection{Existing Drawbacks and Our Motivation }

Reviewing the existing frameworks, we find that they mainly adopt operator-centric optimization, i.e., they optimize the computation graphs by replacing the operators with some fused/split ones to improve the computation inefficiency. We argue such optimization strategies are not globally sufficient and cannot save the performance overheads caused by the spoiled data locality and hardware heterogeneity. Therefore, we turn to develop a new framework, \sysname, and implement its optimization from a dataflow perspective. Compared with the existing works, \sysname enjoys the following advantages.

(1) \sysname possesses a rich operator library and each operator supports multiple dataflow patterns. Given an under-optimized computation graph, \sysname can automatically choose the dataflow pattern for each operator to improve the inference performance. Compared with the operatior-centric optimization adopted by prior works, \sysname's dataflow-centric approach conducts more in-depth optimization and facilitates the continuous maintenance.


(2) \sysname preserves the data locality with the operator linking technique. During the inference computation, \sysname can automatically derive the optimal dataflow pattern for the operators, so as to make a match between (a) the writing order of intermediate parameters output from the operators and (b) the reading order of the subsequent operators. Thus, \sysname can avoid the costly cache misses throughout the whole inference process.

(3) \sysname fully leverages the computation/memory resource with the DSP-aware operator split technique. \sysname partitions the input tensors across multiple DSP units and further splits them to fit into the private memory of every single DSP unit. Thus, it improves the inference parallelism and reduces the overheads of data fetch.


\section{Architecture Overview}
\begin{figure}[!t]
	\centering
	\includegraphics[width=0.45\textwidth]{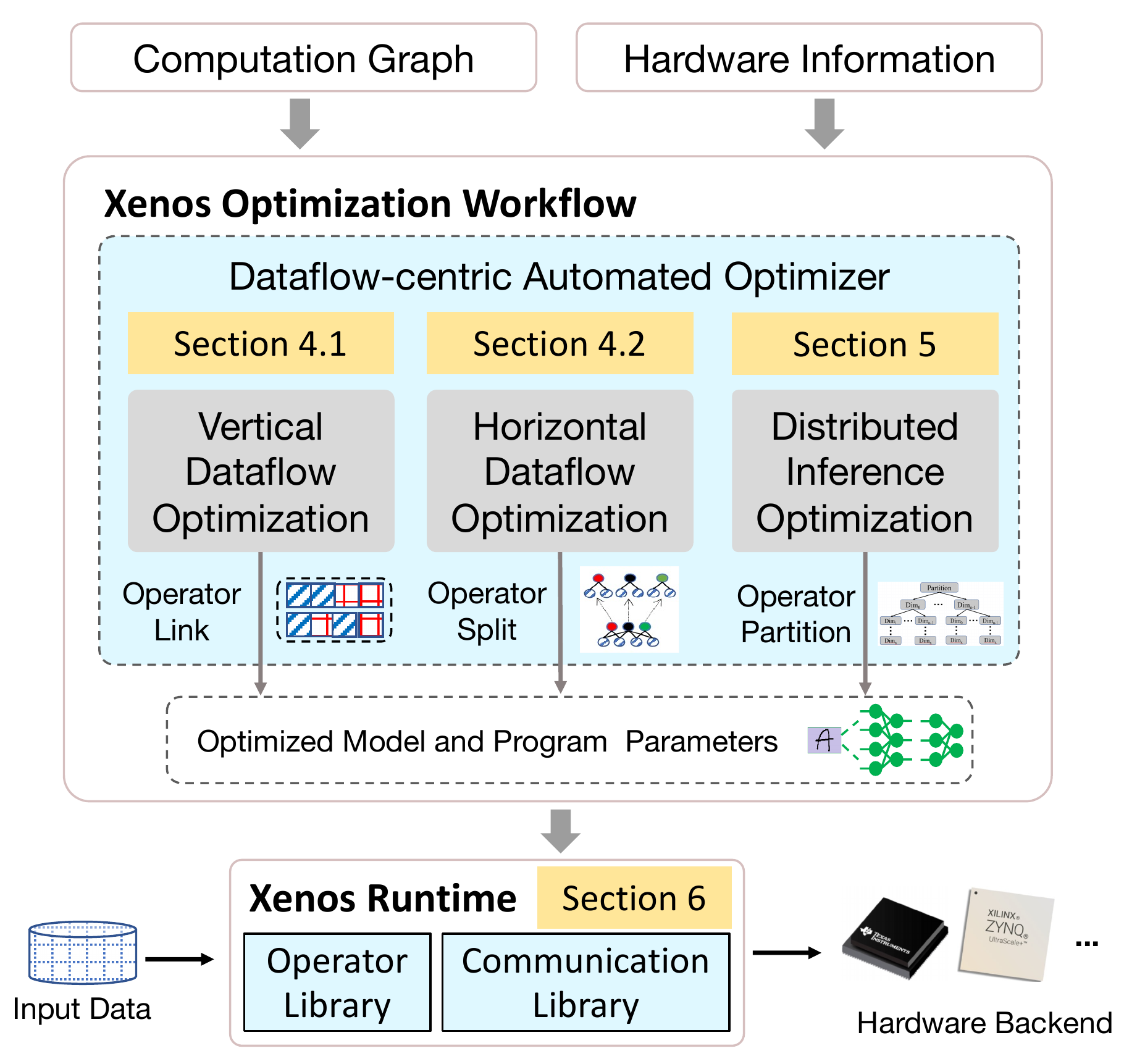}
	\caption{The architecture of \sysname}
	\label{arch}
	\vspace{-0.5cm}
\end{figure}
\Cref{arch} illustrates the architecture of \sysname. Given an under-optimized model, \sysname conducts an automatic optimization workflow based on the hardware information of the edge device. Afterwards, \sysname outputs an optimized model equivalent to the original model. During runtime, both the input data and the optimized model are fed into \sysname Runtime. \sysname Runtime employs multiple DSP units to run the inference task atop its high-performance operator library and communication library.



As in typical frameworks (TASO and PET), \sysname' optimization workflow conducts operator fusion during the preprocessing stage to reduce some inefficient computation (not shown in Figure~\ref{arch}). However, such basic optimization is insufficient for high-performance inference. Therefore, \sysname provides two key techniques, namely, operator linking and DSP-aware operator split to implement vertical and horizontal dataflow optimization, respectively (\S\ref{sec-design}). Besides, \sysname also supports distributed inference (\S\ref{sec-distributed-inference}): it can employ multiple edge devices to share the workload and jointly conduct one consuming inference task.

\section{Design of Xenos}
\label{sec-design}


In this section, we explain the design details of \sysname on the single edge device. Specifically, \sysname optimizes the inter-operator dataflow with operation linking technique (\S\ref{sec-dr-strategy}) and optimizes the dataflow across multiple DSP units with DSP-aware operator split (DOS) technique (\S\ref{sec-hap-strategy}). All these optimization can be automatically executed by \sysname (\S\ref{sec-auto}).

\subsection{Vertical: Operator Linking}
\label{sec-dr-strategy}



\begin{figure}[!t]
	\centering
	\includegraphics[width=0.45\textwidth]{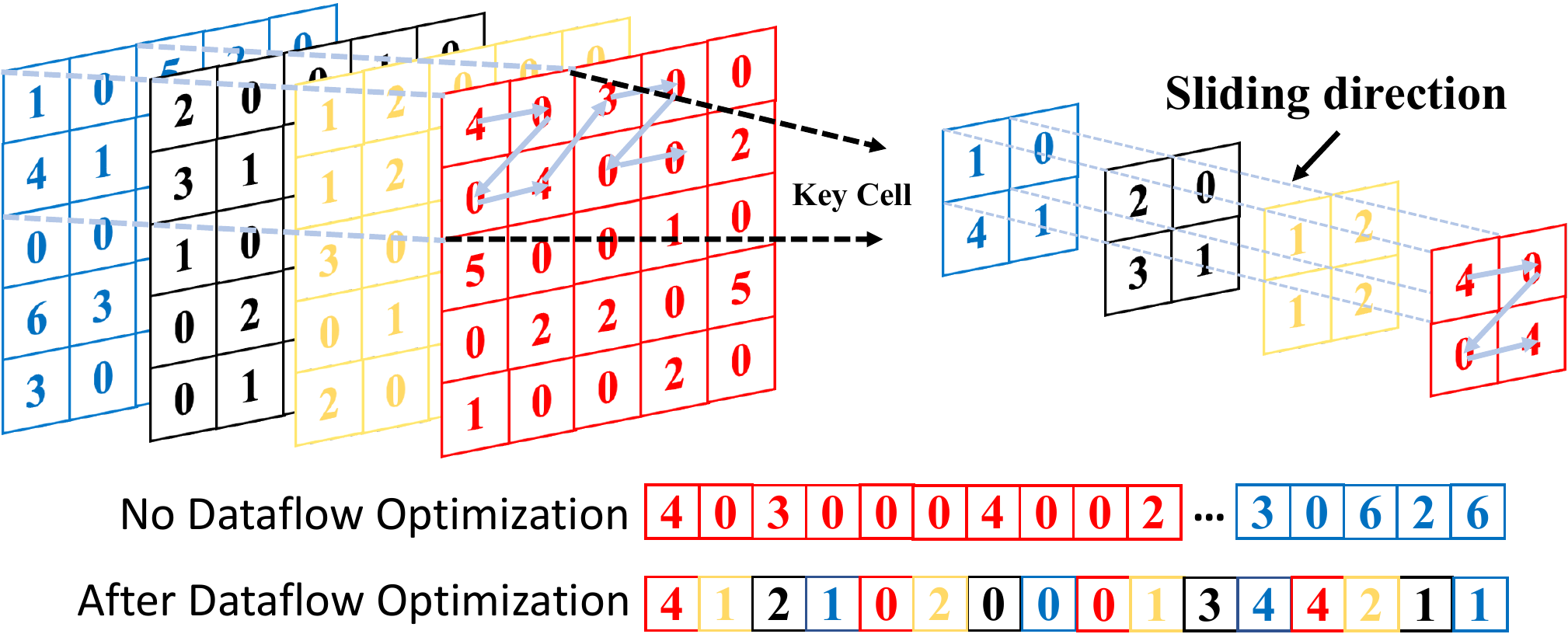}
	\caption{Operator linking to optimize dataflow}
	\label{fusionRestructure}
	\vspace{-0.5cm}
\end{figure}

During the inference, \sysname runtime, after finishing the computation task of one operator, should output the feature map to the shared memory in the desirable layout, so that these data can be read sequentially with good locality during the inference computation with subsequent operators. Otherwise, the data access would suffer from serious overheads due to cache misses. \sysname incorporates the operator linking technique to preserve the data locality. We exemplify the technique  with a linked operator ({\tt Conv1x1} + {\tt AvgPooling2x2}) in Figure~\ref{fusionRestructure}.

In Figure~\ref{fusionRestructure}, the input feature map is generated by the previous operator which has 4 output channels, so the input feature map consists of four matrices (marked with different colors). Without dataflow optimization, the four matrices are placed into the memory one by one in a row-based manner, with the red one placed first, and the blue one placed last. However, during the inference computation, the fused operator needs to read $1\times1$ feature map from each matrix every time, and then computes the average on every $2\times2$ square after the convolution. As magnified on the right side of Figure~\ref{fusionRestructure}, the dataflow goes through the four matrices every time ({\tt Conv1x1} computation). On each matrix, it follows the zigzag pattern ({\tt AvgPooling} computation), leading to the restructured dataflow. Comparing the unoptimized and optimized dataflow, we can see that the unoptimized dataflow suffers from compulsory cache misses for each data access. By contrast, the optimized dataflow completely matches the write/read order during the inference, so it maximizes the data locality and data can be fetched in higher efficiency.

To maintain desirable data locality, the previous operator is required to be aware of the data access pattern of the subsequent operator when it outputs the feature map. With such awareness, \sysname can modify the data layout following the access pattern, instead of simply outputting the matrices one by one. \sysname uses the operator linking technique to attain this, and we describe the details below. 

Before running the inference model, \sysname scans the computation graph and generates the metadata to describe the dataflows in the computation graph. Then, \sysname analyzes the metadata and identify the specific patterns (i.e., a sequence of adjacent operators) that can spoil data locality. After finding such inefficient patterns, \sysname will modify the metadata to change the dataflow between these adjacent operators. The metadata is fed into the inference engine. During runtime, the inference engine can know from the metadata the data access pattern of the subsequent operator. Therefore, it can write the feature map according to the optimized dataflow with data locality preserved (see Figure~\ref{fusionRestructure}).

Notably, the operator linking technique can also incur data redundancy when conducting the computation of standard convolution, because it replicates some parameters of the feature map to avoid the subsequent operator from looking back. However, the memory sacrifice proves to be worthwhile, because the performance benefit brought by the restructuring outweighs the additional memory cost and the inference workflow is effectively accelerated (shown in \S\ref{sec-exp-at}).

\subsection{Horizontal: DSP-Aware Operator Split}
\label{sec-hap-strategy}

\sysname incorporates DOS to partition the inference workload for higher parallelism. DOS focuses on two aspects. First, it needs to partition the feature map across multiple DSP units, so that the multiple DSP units can share the inference workload. Second, it needs to split the operator parameters into the memory hierarchy of the edge device, so that the parameter fetch can be more efficient.

\subsubsection{Partition Feature Map} 
\label{sec-ha-strategy}



\sysname partitions the feature map~\footnote{As a special case, when the feature map is too large to be held by the shared memory, \sysname will first slice the feature map as a preprocessing step, and DOS continues to partition the sliced feature map.} in three dimensions, namely, the input feature map height ($inH$), the input feature map width ($inW$), and the output feature map channel ($outC$). We dismiss the input-channel-based ($inC$) partition since the $inC$-based partition  performs extra reduction and introduces more computation overheads. Considering that the feature maps are stored in the shared memory with size of $4$ MB, which is far beyond the input channel size in typical models, we usually do not have to partition along input channel. 

\sysname prioritizes $outC$-based partition due to its less complexity: \sysname simply distributes the kernel parameters to different DSP units, and these kernel parameters will be placed into the L2 memory of DSP units. All DSP units can access the feature map located in the shared memory. On the other hand, $inH$-based scheme and $inW$-based scheme partition the remaining feature map after $outC$-based partition, and they usually require special handling of the boundary rows/columns. Only if the kernels cannot be evenly distributed across DSP units, DOS will seek further partition by $inH$/$inW$. If imbalance still exists after the triple partition, DOS will randomly assign the remaining feature map (workload) to different DSP units.


While the operator partition procedure distributes the inference workload across multiple DSP units, the single DSP unit may still fail to conduct the inference work efficiently, because the operator parameters assigned to it is too large to fit into the L2 memory (e.g. the CNN model has very large-sized kernels). To address that, \sysname needs to do further split of the operator parameters according to the memory resource of the DSP unit. We explain the split of operator parameters next.

\subsubsection{Split Operator Parameters}
\label{sec-ma-split}

\sysname splits the large-sized operator parameters into smaller chunks so that they can be placed into the private L2 memory of the DSP unit. Thus, the parameter fetch can become more efficient and the inference time can be reduced. 

\sysname follows a certain priority for each dimension when performing parameter splitting, to guarantee that minimum computation overhead is introduced after splitting.
Taking the popular CNN model as an example, which have four dimensions for parameters, i.e., output channel ($K$), input channel ($C$), kernel height ($R$), and kernel width ($S$).
Splitting at $K$ dimension will not introduce any extra computation, while splitting at the other three dimensions requires additional reduction operation to aggregate the results on these dimensions, which introduces extra computation overhead.
Thus, \sysname will first try to split the parameters at $K$ dimension, and then $C$, $R$ and $S$ if only splitting $K$ dimension is not enough to fit the parameter in L2 memory.

Equation~\ref{f4} gives an example of output-channel-based split. Since the large-sized parameters ($W$ and $B$) can not be put into the L2 memory, \sysname performs fine-grained split of the operator:
$W$ is split into $W_1$ and $W_2$, and $B$ is split into $B_1$ and $B_2$. 
After that, parameters can be distributed into the L2 memory of two different DSP units. 
Equation~$y_0$ and $y_1$ can be jointly executed on two DSP units in parallel, or on one DSP unit one by one.
The output $y_{1}(x_{i})$ and $y_{2}(x_{i})$ are automatically joined together afterwards, without performing any data layout transformation operators. 
Other types of splitting work in a similar way.
\begin{align} 
\begin{aligned}
\label{f4}
y(x_{i}) &= Wx_{i} + B \\
&\Downarrow \\
y_{1}(x_{i}) &= W_{1}x_{i} + B_{1} \\
y_{2}(x_{i}) &= W_{2}x_{i} + B_{2}
\end{aligned}
\end{align}



\subsection{Exemplar Optimization } 
\label{sec-example-oec}
\begin{figure}[h]
	\centering
	\includegraphics[width=0.48\textwidth]{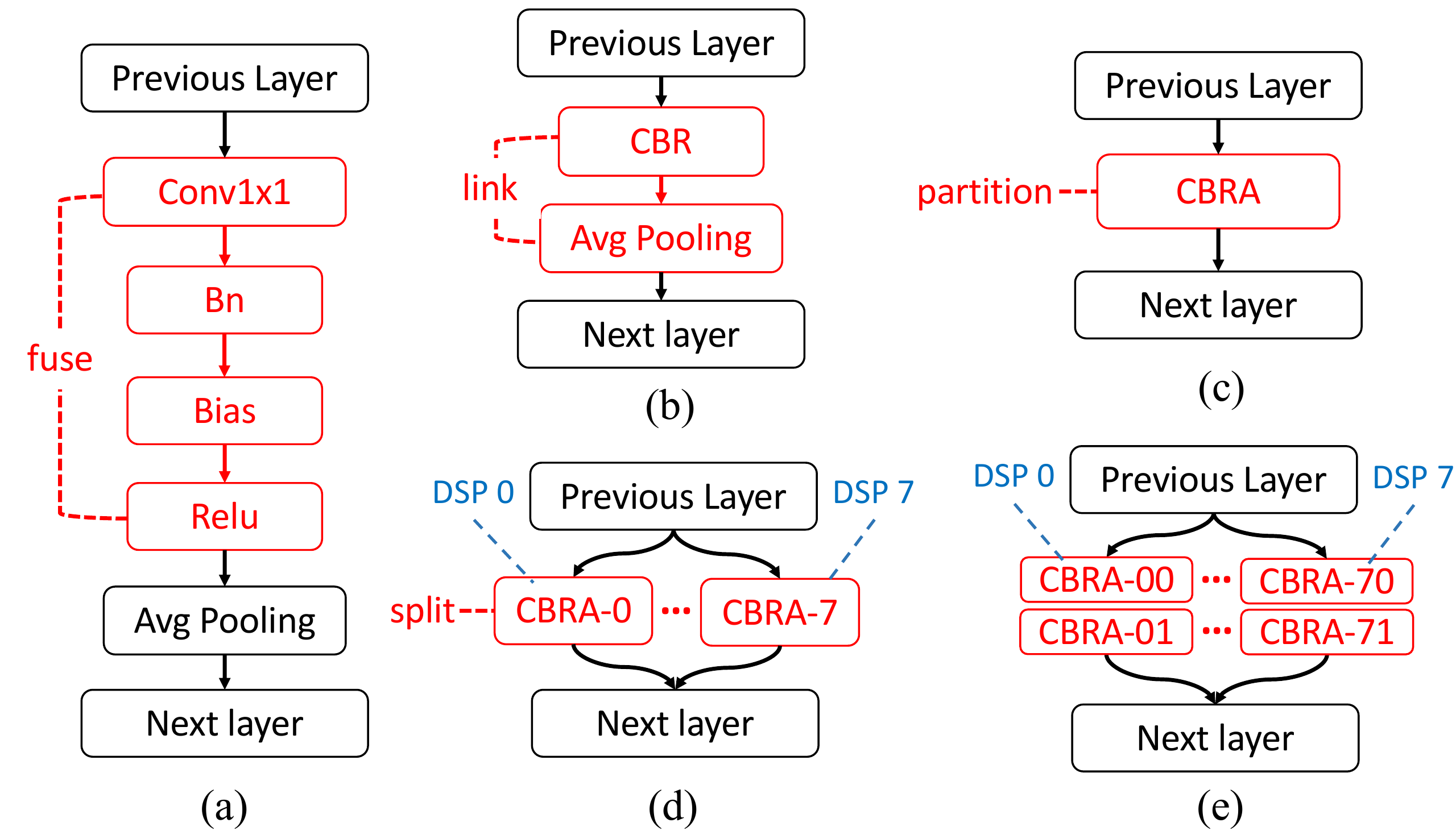}
	\caption{Illustration of \sysname's optimization }
	\label{opfusion}
\end{figure}

\Cref{opfusion} gives an example of how \sysname optimizes one part of the computation graph of MobileNet~\cite{howard2017mobilenets}. 
\sysname first performs typical operator fusion on operator {\tt Conv1x1} (convolution), {\tt Bn} (batch normalization), {\tt Bias}, and {\tt Relu}, and generates a fused operator {\tt CBR} (Figure 5(a)), then links {\tt CBR} and {\tt AvgPooling} together to form the operator {\tt CBRA} (Figure 5(b)).
Next, \sysname uses DOS technique to partition the feature map across DSP units (Figure 5(d)). After that, the operator parameters are still too large to fit each DSP unit's L2 memory, so \sysname continues to split the operator parameters into smaller chunks (Figure 5(e)) and finalizes the optimization.




\subsection{Automatic Optimization}
\label{sec-auto}
\sysname has been equipped with the automatic procedure to conduct both vertical and horizontal optimization. Without any manually tuning effort, \sysname can identify the proper patterns in the computation graph (Table~\ref{AutomaticPI}), and replace the under-optimized dataflows with an optimized version. The automation brings great convenience for developers to deploy the optimized model. We have measured the time cost of the automatic optimization on multiple typical inference models (Table~\ref{AutomaticTC}), which takes \SI{0.11}{\second}--\SI{0.91}{\second} to transform the under-optimized model to the optimized one.

\begin{table}[h]
\renewcommand\arraystretch{1.2}
	\centering
	\caption{Automatic pattern identification}
	\label{AutomaticPI}
\scriptsize
	\begin{tabular}{ccc}
		\hline
		Linking & Examples\\
		\hline
                        &{\tt1.Conv 3$\times$3 -> Conv1$\times$1}  \\
                        &{\tt2.Conv 5$\times$5 -> Conv1$\times$1}  \\
  {\tt ConvX -> ConvY}  &{\tt3.Conv 7$\times$7 -> Conv1$\times$1}  \\
                        &{\tt4.Conv 3$\times$3 -> Conv3$\times$3}  \\
                        &{\tt...}  \\
		\hline
                        &{\tt1.Conv3$\times$3 -> Conv1$\times$1 -> AvgPooling}  \\
   {\tt ConvX -> ConvY -> ZPooling} &{\tt2.Conv3$\times$3 -> Conv1$\times$1 -> MaxPooling}  \\
   {\tt ConvX -> ZPooling -> ConvY} &{\tt3.Conv1$\times$1 -> AvgPooling -> Conv3$\times$3}  \\
                                    &{\tt4.Conv1$\times$1 -> MaxPooling -> Conv3$\times$3}  \\
                        &{\tt...}  \\
		\hline
		&\\
  {\tt \(\operatorname{\tt ConvX  \tt -> } \left\{\begin{array}{c}\cdots -> \text { ConvY} \\ \text { ConvZ}\end{array}\right.\)} &{\tt Shortcut Connection}\cite{cvpr16-resnet}  \\
        &\\
		\hline
  {\tt MatmulX -> MatmulY } &{\tt MatA * MatB -> MatC * MatD}  \\
		\hline
	\end{tabular} 
 	\vspace{-0.1cm}
\end{table}

\begin{table*}[h]
\renewcommand\arraystretch{1.2}
	\centering
	\caption{Automatic Time Cost}
	\label{AutomaticTC}
\footnotesize
	\begin{tabular}{cccccccc}
		\hline
		&MobileNet & SqueezeNet&  ShuffleNet  &ResNet18& CentreNet& LSTM & Bert-S\\
		\hline
		Time cost (s) &$0.11$ &$0.14$ &$0.36$ &$0.24$ &$0.18$ &$0.64$ &$0.91$  \\
		\hline

	\end{tabular} 
 	\vspace{-0.1cm}
\end{table*}

\section{Distributed Inference} 
\label{sec-distributed-inference}

We envision that the inference workload can soon go beyond the capacity of single edge devices. Recent models, such as ResNet-101 (60.2M)~\cite{cvpr16-resnet}, Bert (340M$\sim$481000M)~\cite{devlin2018bert} and GPT-3\cite{radford2018improving}(175000M), can hardly be used for single-device inference, making distributed inference become a necessity.
Therefore, we extend \sysname to leverage multiple devices for joint inference computation, and we call the distributed version of \sysname as \dsysname. 





\algrenewcommand\algorithmicrequire{\textbf{Input:}}
\algrenewcommand\algorithmicensure{\textbf{Output:}}
\begin{algorithm}[!htbp]
\small
\makeatletter
\NewDocumentCommand{\LeftComment}{s m}{%
  \Statex \IfBooleanF{#1}{\hspace*{\ALG@thistlm}}\(\triangleright\) #2}
\makeatother

\algdef{SE}[EVENT]{Event}{EndEvent}[1]{\textbf{upon}\ \textsc{\small #1}\ \algorithmicdo}{\algorithmicend\ \textbf{event}}%
\algtext*{EndEvent}
\caption{Enumerating Partition Schemes}
\label{algo:enumerate}

  \begin{algorithmic}[1]
  \Require  $dset$--{The dimension set to be partitioned}
      \State $bestShm$, $bestTime$ = $\varnothing$, $+\infty$
      \For {$shm$ $\in$ \textsc{permutation}(dset)}
        \State $execTime = $ \textsc{Profiling}($shm$)
        \If {$execTime < bestTime$}
            \State $bestShm$, $bestTime$ = $shm$, $execTime$
        \EndIf
      \EndFor
      \State \Return $bestShm$
      
  \end{algorithmic}
\end{algorithm}

\dsysname supports two methods to synchronize the parameters across multiple edge devices, namely, parameter server (PS)-based synchronization~\cite{osdi14_ps} and ring all-reduce synchronization~\cite{ring-allreduce}. The key idea of \dsysname is to incorporate both model-parallel computation and  parameter synchronization among multiple devices, to cut down the overall inference time. 
However, we note that, compared with \sysname, \dsysname may not achieve optimal inference time if it still follows the priorities described in \S\ref{sec-ha-strategy} to partition the feature map. 
Taking convolution operator as an example, recall that convolution's DOS prioritizes \emph{outC}-based partition over \emph{inH}-/\emph{inW}-based partition. 
The reason is that \emph{outC}-based partition can fully leverage the shared memory across different DSP units and avoid padding overheads due to  \emph{inH}-/\emph{inW}-based partition. 
However, when it comes to \dsysname, different edge devices do not share memory, and it can hardly tell whether or not \emph{outC}-based partition can outperform \emph{inH}-/\emph{inW}-based partition. 
To address this problem, \dsysname uses a tree based traversal algorithm to determine the partition strategies.

\begin{figure}[!t]
    \centering
    \includegraphics[width=0.45\textwidth]{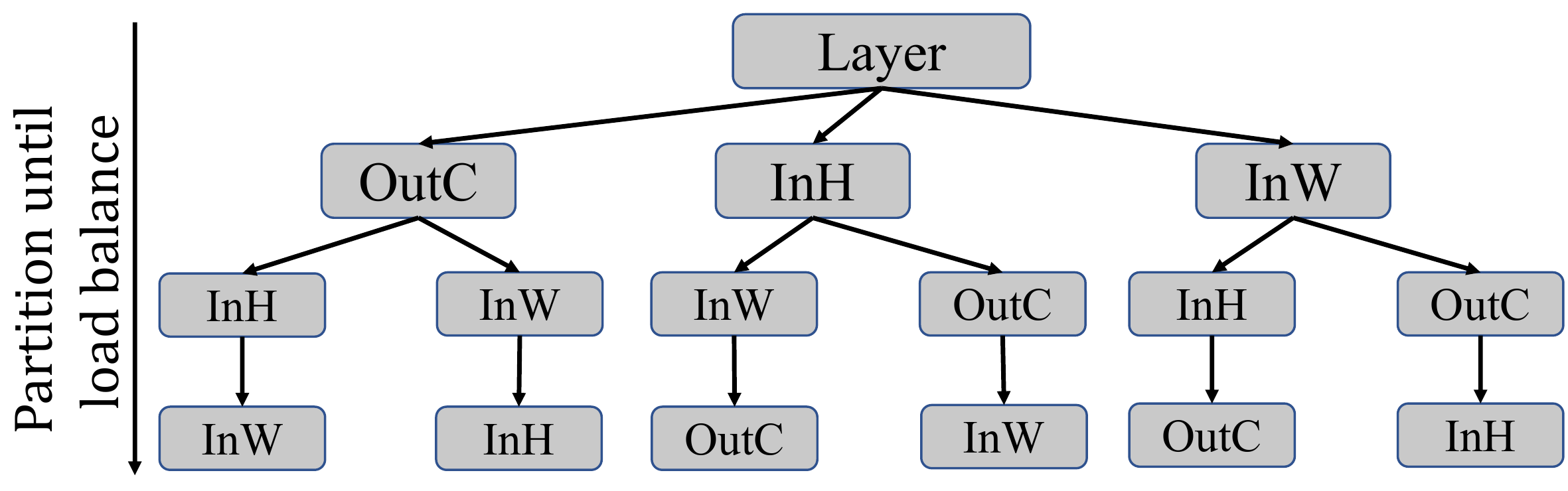}
    \caption{Enumerating partition schemes in \dsysname}
    \label{fig-branch-partition}
\end{figure}

The general enumeration algorithm is described in Algorithm~\ref{algo:enumerate}. \dsysname enumerates every possible combination of  partition schemes towards dimensions appeared in feature maps or operator parameters involved in the inference, which can reach $d!$ different combinations at most if there are totally $d$ dimensions.
For example, there are three dimensions in matrix multiplication ($m$, $n$, and $k$, where $k$ is shared by both feature map and parameter) and seven dimensions in convolution ($batch$, $inC$, $inH$, $inW$, $outC$, $r$ and $s$, where $inC$ is shared by both feature map and parameter). However, as we discussed in \S\ref{sec-ha-strategy} that \sysname only partitions along certain dimensions, \dsysname also chooses these dimensions to partition because we empirically notice that the other dimension-based partition (e.g. $inC$-based partition for convolution operator) cannot achieve better performance in the distributed setting either.
Therefore, \dsysname still focuses on the partition schemes along \emph{inH}, \emph{inW} and \emph{outC} when partitioning a convolution, as shown in Figure~\ref{fig-branch-partition}.
Then \dsysname chooses the best partition strategy among them according to the profiling result of their execution time.
Due to such search process is one-off for each model when deploying on a specific platform, the search cost is acceptable.

\section{Implementation}
\label{sec-imple}
To use \sysname, users need to provide a computation graph for the inference model, along with the hardware-specific information for a edge device, e.g., memory hierarchy, number of DSP units, etc., so that \sysname can perform hardware-adaptive optimizations. The optimized model will be generated as an executable supported by \sysname runtime, including an efficient operator library, and the middlewares for memory management and scalable communication. Currently, \sysname can support different edge devices including Multi-Core DSP hardware~\cite{tms}, Xilinx U-series hardware~\cite{useries}, Xilinx Zynq-series hardware~\cite{zcu}, etc.



\subsection{{Operator Library}}

Existing frameworks tend to offer API interfaces with high-level languages (e.g. Python) and mask the low-level implementation details to programmers. However, while working with edge hardware platforms, programmers are required to use the low-level operators and programming languages (e.g. C/C++ and Assembly languages) to implement their own functions. 


Our goal is to provide ordinary staff with a rich operator library, which is efficient and can be easily used to implement their high-level functions on these edge hardware platforms. Therefore, we begin by implementing with low-level languages (C/C++ and Assembly) various types of operators (e.g. standard convolution with different KSize and different Stride, depthwise separable convolution, MaxPooling/Global Pooling, Concat, etc.). 
Meanwhile, we keeps iterating over the operator library for usability and compatibility. The basic operators provided by \sysname can be summarized as Table~\ref{xenos-op}.  Nowadays different ML frameworks tend to provide hundreds of operators. The cost of operator development of maintenance can be expensive to support their operator-centric optimization strategies, because adding one optimization strategies usually needs inventing more operators. By contrast, \sysname' operators support customized dataflow between operators. To implement its vertical and horizontal optimization, \sysname only needs to customize the inter-operator dataflows, instead of adding more operators, which saves the maintenance cost and can be easily extended in the future development.


\begin{table}[t]
\renewcommand\arraystretch{1.2}
\small
\centering
\caption{{Operators and Optimization}}
\label{xenos-op}
	\begin{tabular}{cc} 
		\hline
		Operator &Description  \\
        \hline
        {\texttt{x.add}}           &{Element-wise Addition}              \\
		{\texttt{x.mul}}           &{Element-wise Multiplication}        \\
		{\texttt{x.mac}}           &{Multiply Accumulate}        \\
		{\texttt{x.conv}}          &{Convolution (kernel size, stride, padding, etc)}\\          
		{\texttt{x.matmul}}        &{Matrix Multiplication}               \\
		{\texttt{x.gampool}}       &{Global / Average / Max Pooling}      \\
		{\texttt{x.transpose}}     &{Matrix Transpose}                    \\
		{\texttt{x.concat}}        &{Concatenation of Multiple Tensors}   \\
		{\texttt{x.split}}         &{Split a Tensor into Multiple Tensors}\\	
		{\texttt{x.cbr}}         &{Fused {\tt Conv-Bn-Relu} operator}\\		
		{\texttt{x.cbrm}}         &{Linked {\tt CBR-MaxPooling} operator}\\		
		{\texttt{x.cbra}}         &{Linked {\tt CBR-AvgPooling} operator}\\		
		\hline
	\end{tabular} 
	\vspace{-0.5cm}
\end{table}

\subsection{Scalable Communication Middleware}
Image data usually requires a series of preprocessing operations before inference. Normally, the preprocessing module and the inference module are not executed on the same device. Therefore, we need to build a bridge between them. For the sake of modularity and scalability, we encapsulate the communication primitive into an independent middleware. The middleware is compatible with both the SRIO protocol (for embedded applications) technology and the conventional Ethernet protocol. Meanwhile, our communication middleware is integrated with pipeline and batch transmission mechanisms, to achieve high throughput performance. Additionally, we also customized the efficient packing/unpacking function for the sake of lower real-time latency.



\section{EXPERIMENTAL EVALUATION}
\label{sec-exp}
We aim to answer the following questions during evaluation:

(1) How much acceleration can \sysname bring to the model inference workflow?

(2) What speedup can \sysname bring to typical operators?

(3) How much time does \sysname cost to complete the automatic optimization?

(4) How much hardware resource can be  saved duing the inference with \sysname?

(5) How much acceleration can the distributed \sysname (\dsysname) achieve, compared with the single-node version?

\subsection{Experiment Setting }
\textbf{Testbeds:}
We employ two testbeds for evaluation: (1) a Multi-Core DSP device, which is equipped with 2 TMS320C6678-type nodes and a high-speed image collector, which are directly connected via SRIO; (2) the ZCU102-type FPGA device, with code generated by Intel High-Level Synthesis (HLS) Compiler. We use TMS320C6678 and ZCU102 to refer to them for simplicity.  

\textbf{Benchmarks:} We choose 7 typical models as the benchmarks, i.e. MobileNet, SqueezeNet, ShuffleNet, ResNet18, CentreNet, LSTM and Bert.  

\textbf{Baselines:}
 We first conduct an ablation study with \sysname: we compare the complete \sysname solution, which have incorporated both horizontal optimization (HO) and vertical optimization (VO) , with two baselines, from which we show the benefit of \sysname's two strategies.  As shown in Figure~\ref{AT}, one baseline does not involve either HO or VO, which we call it Vanilla baseline. The other baseline only adopts HO, which we call it HO baseline. Then, we compare \sysname with TVM~\cite{osdi18-tvm}, running both of them with the same benchmarks on ZCU102. Besides, we also compare \sysname with a GPU baseline, where we run the same benchmarks with PyTorch on an NVIDIA RTX 3090 GPU.
 

\textbf{Metrics:}
We study the inference time cost and the hardware resource cost in our evaluation. As for the inference time cost, we run each inference workload for 1000 times and report the average value. 
As for the resource cost, we compare the cost of L2, SRAM, and DDR on TMS320C6678; and we compare the cost of DSP~\footnote{DSP (refer to DSP slice) is composed of high-speed multiplier circuits to perform high-speed multiply accumulate operations in FPGA.}, FF~\footnote{FF (Flip Flop) is a storage unit that can only store 1 binary bit and can be used as a memory element for sequential logic circuits.} and LUT~\footnote{LUT (Look-up Table) is the module used to implement the function of the combinational logic circuit.} on ZCU102.



\subsection{Inference Time Comparison}
\label{sec-exp-at}

\begin{figure}[!t]
	\centering
	\subfigure[TMS320C6678]{
		\label{6678AT} 
		\includegraphics[width=0.45\textwidth]{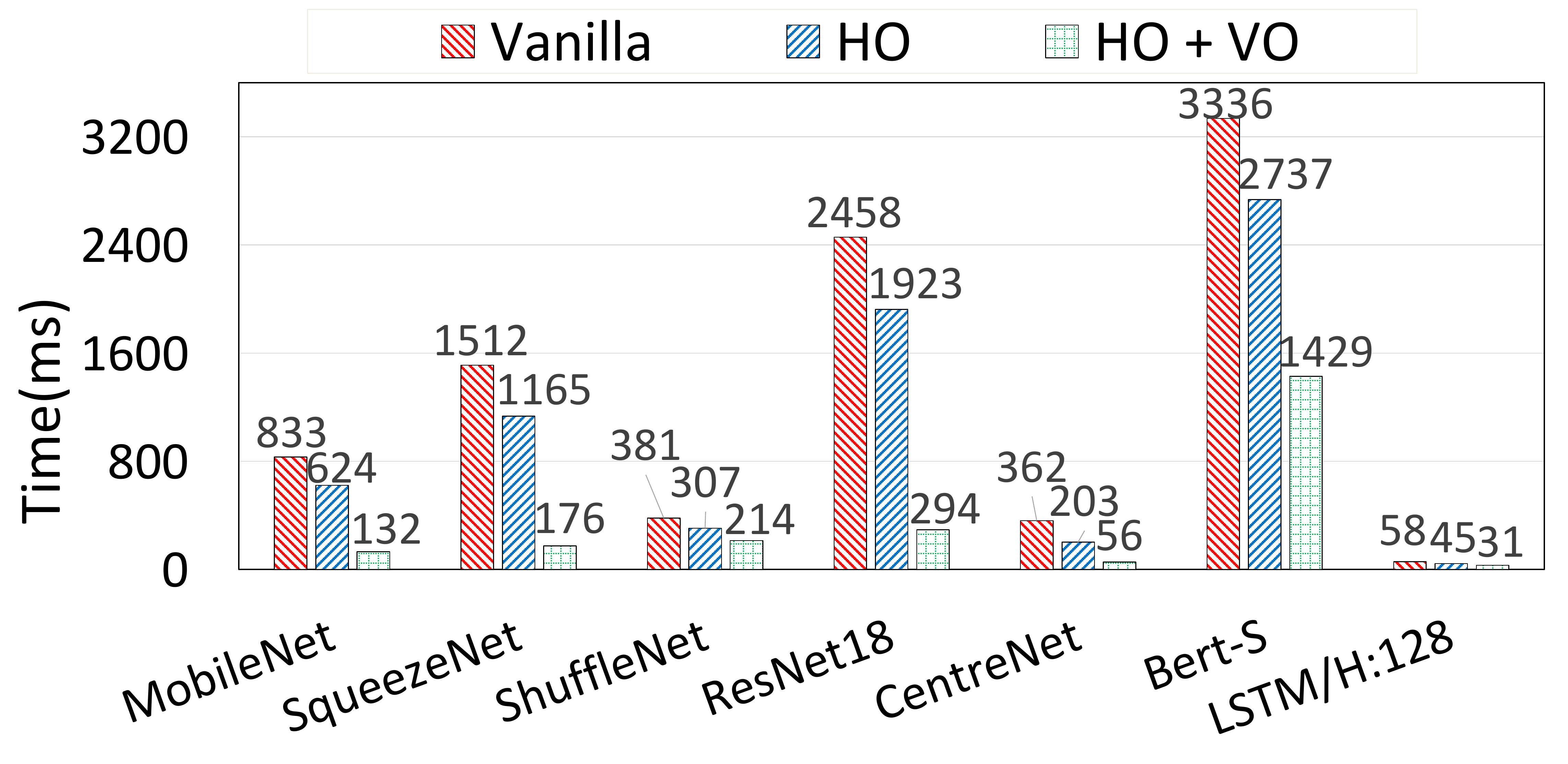}
	}
	\\\vspace{-0.3cm}
	\subfigure[ZCU102]{
	\label{fpgaAT} 
	\includegraphics[width=0.45\textwidth]{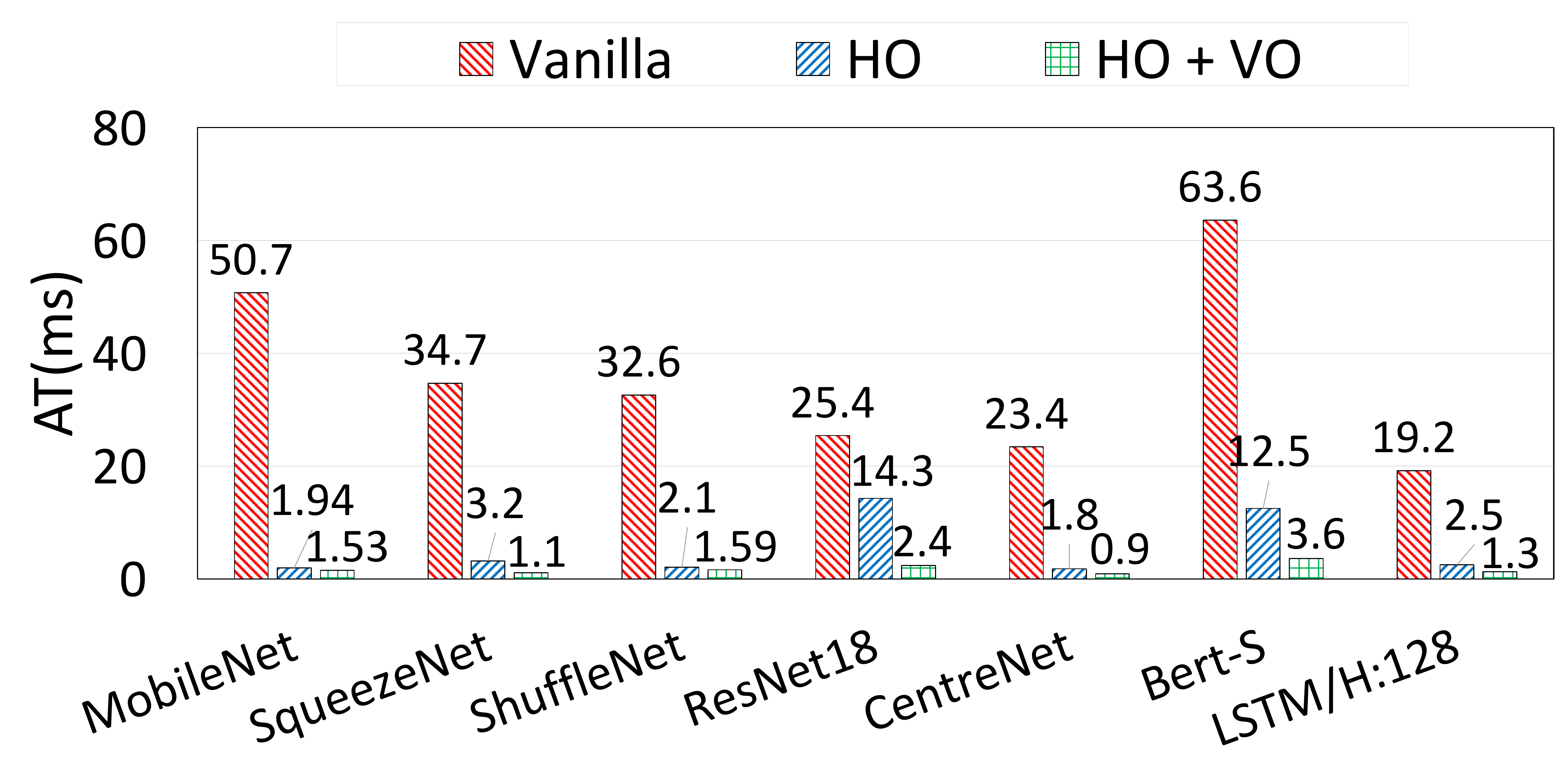}
	}
	\vspace{-0.3cm}
	\caption{Inference time comparison}
	\label{AT} 
	\vspace{-0.5cm}
\end{figure}

\para{TMS320C6678.} In Figure~\ref{6678AT}, compared with Vanilla, HO reduces the inference time by 17.9\%-43.9\%, which demonstrates the acceleration brought by higher computation parallelism. We further evaluate the performance benefit of VO by comparing the HO baseline and the full \sysname solution. We can see that the VO further reduces the inference  time by 30.3\%-84.9\%, which demonstrates the performance improvement brought by the good data locality during the inference.






\para{ZCU102.} Figure~\ref{fpgaAT} also demonstrate the performance benefit of HO and VO on ZCU102. Similarly, compared with the Vanilla baseline, HO can reduce the inference time by 80.4\%-96.2\%. Compared to the HO baseline, VO further reduces the inference time by 21.2\%-83.3\%. 


Comparing Figure~\ref{6678AT} and Figure~\ref{fpgaAT}, we can easily notice that HO contributes more inference time reduction on TMS320C6678, whereas VO contributes more on ZCU102. The reasons are explained in two main aspects. 
\begin{enumerate}[wide, labelwidth=!,nosep, label=(\arabic*)]
    \item VO is more effective on TMS320C6678 than ZCU102. This is because a large number of LUT resources are used on ZCU102 to implement data mapping, therefore, the memory access efficiency has already been very high even without VO. By contrast, TMS320C6678 is not equipped with such a utility, so the memory access efficiency can be seriously damaged when the inference workload breaks the data locality. VO, however, helps to preserve the data locality with dataflow restructuring and becomes the main contributor.
    \item HO is more effective on ZCU102 than TMS320C6678. This is because ZCU102 has much more DSP units than TMS320C6678. While TMS320C6678 only has 8 DSP units, ZCU102 can allocate thousands of DSP units to participate in the computation. Therefore, the management of model partition and parallel execution becomes more essential to the efficiency of ZCU102. Since the Vanilla baseline is not equipped with a proper partition scheme, it fails to exploit the abundant computation resource. In contrast, 
    HO works with the optimized operators and achieves high utilization of DSP computation resource, which can significantly  reduce the inference time. 
\end{enumerate}

\begin{figure}[!t]
	\centering
	\includegraphics[width=0.45\textwidth]{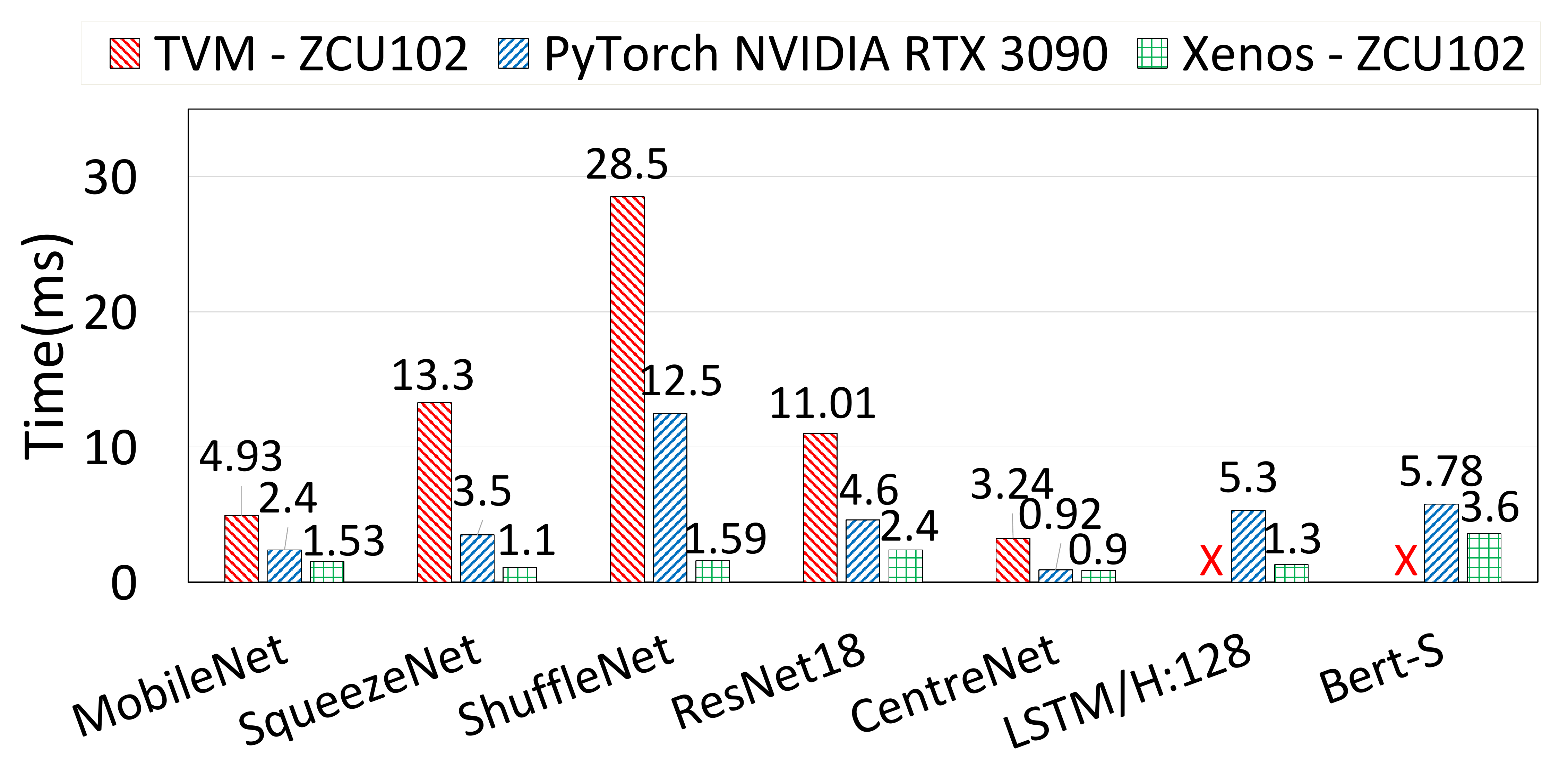}
	\caption{Inference time comparison with TVM and PyTorch}
	\label{TVMGPUXenos}
	\vspace{-0.5cm}
\end{figure}


\para{Comparing with Other Baselines.}  We further compare \sysname with two other baselines. First, we run TVM on ZCU102 with the same inference models\cite{tvmdeploy}. Second, we also report the inference performance in a GPU environment: we use PyTorch and run the same models with an NVIDIA RTX 3090 GPU. Figure~\ref{TVMGPUXenos} shows \sysname also significantly outperforms the other two baselines. To be more specific, \sysname achieves 1.02$\times$--1.87$\times$ speedup compared with the GPU baseline, and it achieves 3.22$\times$--17.92$\times$ speedup compared with the TVM baseline.  Across all the benchmark models, except LSTM and Bert-S~\footnote{TVM cannot run LSTM/Bert-S on edge hardware, because TVM is relies on the development kit provided by Xilinx to run the inference models, but Xilinx's development kit does not support running LSTM/Bert-S on ZCU102}, TVM falls far behind \sysname running on the same hardware, because TVM fails to fully exploit the hardware information during the inference process.  


\subsection{Micro-Benchmark on Typical Operators}
\vspace{-0.3cm}
\label{sec-op-optim}

\begin{table}[h]
\renewcommand\arraystretch{1.2}
	\centering
	\caption{Speedup for typical operators. {\tt CBR} is the abbrevation for {\tt Conv-Bn-Relu}.}
	\label{OEC}
\footnotesize
	\begin{tabular}{ccc}
		\hline
		Operators & \sysname Optimization&  Speedup\\
		\hline
		{\tt CBR-MaxPooling} &  & \\
		$224\times224\times24$ & Operator Linking& $3.3\times$  \\
		$3\times3\times3\times224$ &  & \\
		\hline
		{\tt CBR-AvgPooling} &  & \\
		$7\times7\times1024$ & Operator Linking& $2.3\times$  \\
		$1\times1\times1024\times1024$ &  & \\
		\hline
		{\tt FullyConnected} &  & \\
		$1\times1\times1536$ & Operator Split& $2.25\times$  \\
		$1\times1\times1536\times1000$& &   \\
		\hline
		{\tt CBR} &  & \\
		$112\times112\times32$ & Operator Split& $2.6\times$  \\
		$1\times1\times32\times64$& &   \\
		\hline
	\end{tabular} 
 	\vspace{-0.1cm}
\end{table}

\begin{table}[h]
\renewcommand\arraystretch{1.2}
	\centering
	\caption{Speedup for typical operators. {\tt CBR} is the abbrevation for {\tt Conv-Bn-Relu}.}
	\label{OEC}
\footnotesize
	\begin{tabular}{ccc}
		\hline
		Operators &  Optimization&  Speedup\\
		\hline
		{\tt FullyConnected} &  & \\
		$1\times1\times1536$ & Operator Split& $2.25\times$  \\
		$1\times1\times1536\times1000$& &   \\
		\hline
		{\tt CBR} &  & \\
		$112\times112\times32$ & Operator Split& $2.6\times$  \\
		$1\times1\times32\times64$& &   \\
		\hline
		{\tt FullyConnected} &  & \\
		$1\times1\times1536$ & - & $1\times$  \\
		$1\times1\times1536\times1000$& &   \\
		\hline
		{\tt CBR} &  & \\
		$112\times112\times32$ & - & $1\times$  \\
		$1\times1\times32\times64$& &   \\
		\hline
	\end{tabular} 
 	\vspace{-0.1cm}
\end{table}



We solicit typical operators, and study \sysname's acceleration for them. We run a micro-benchmark on TMS320C6678 with typical operators in CNN, and Table~\ref{OEC} shows the speedup brought by \sysname.



\subsection{Automatic Time Cost}
We would like to evaluate how long it takes for \sysname to complete the automatic optimization of an inference model. Table~\ref{AutomaticTC} lists the time cost to optimize the typical inference models. Most optimization can be completed in 0.1--0.2 seconds. For complex models such as Bert-S, \sysname' automatic optimization costs longer time but still less than 1 second.

\subsection{Resource Cost Comparison}

\begin{figure*}[t]
\vspace{-0.2cm}
	\centering
	\subfigure[MobileNet L2 Cost]{
		\label{Ml2} 
		\includegraphics[width=2.1in]{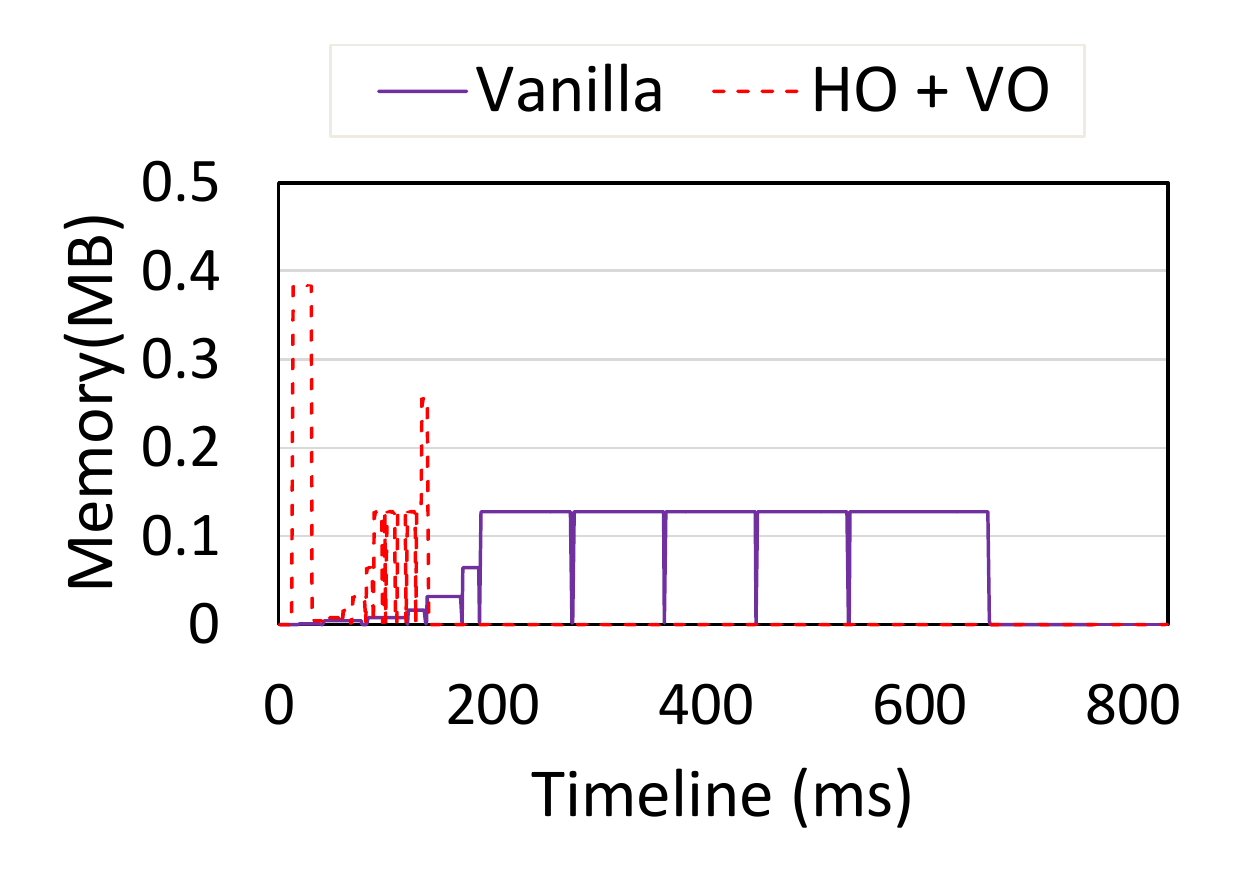}
	}
	\subfigure[MobileNet SRAM Cost]{
    	\label{Msr} 
    	\includegraphics[width=2.1in]{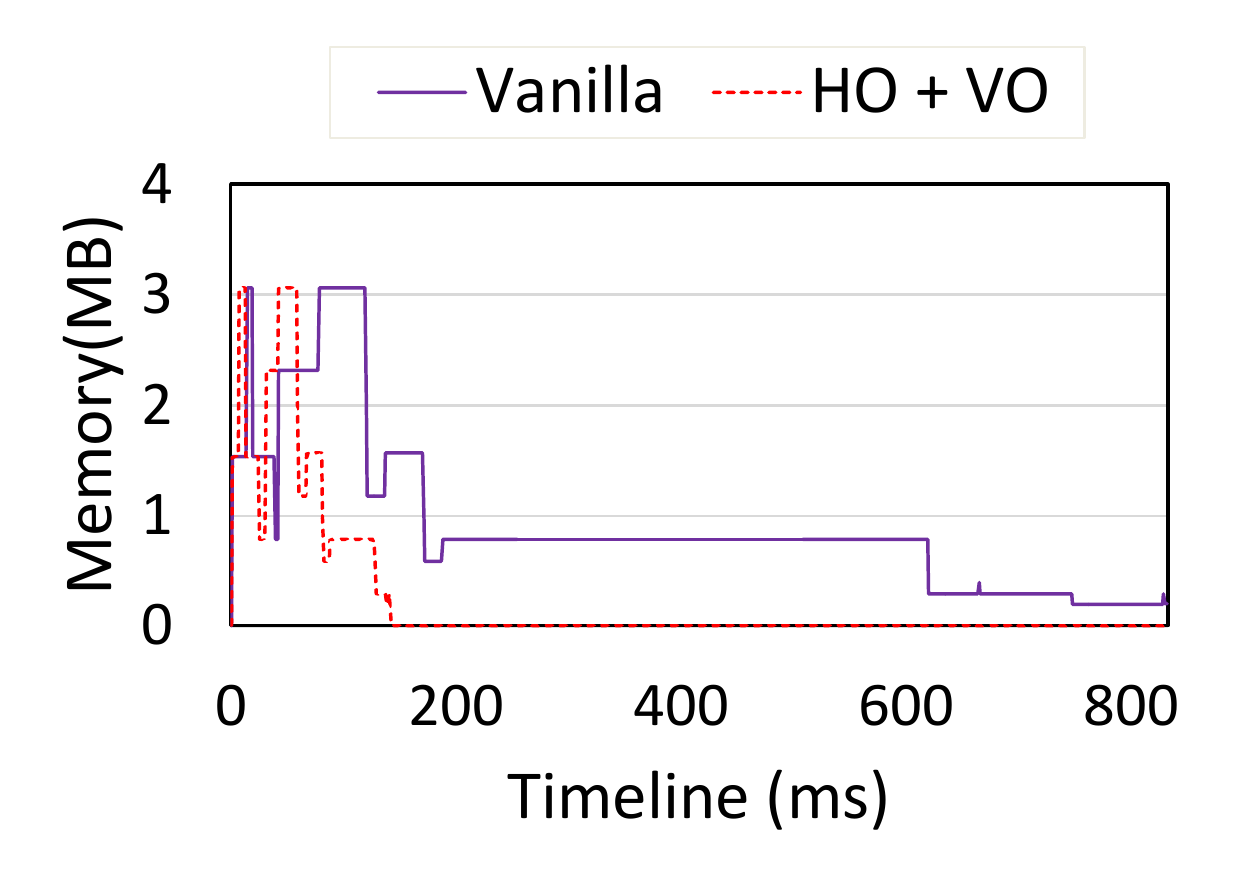}
	}
	\subfigure[MobileNet DDR Cost]{
		\label{Mddr} 
		\includegraphics[width=2.1in]{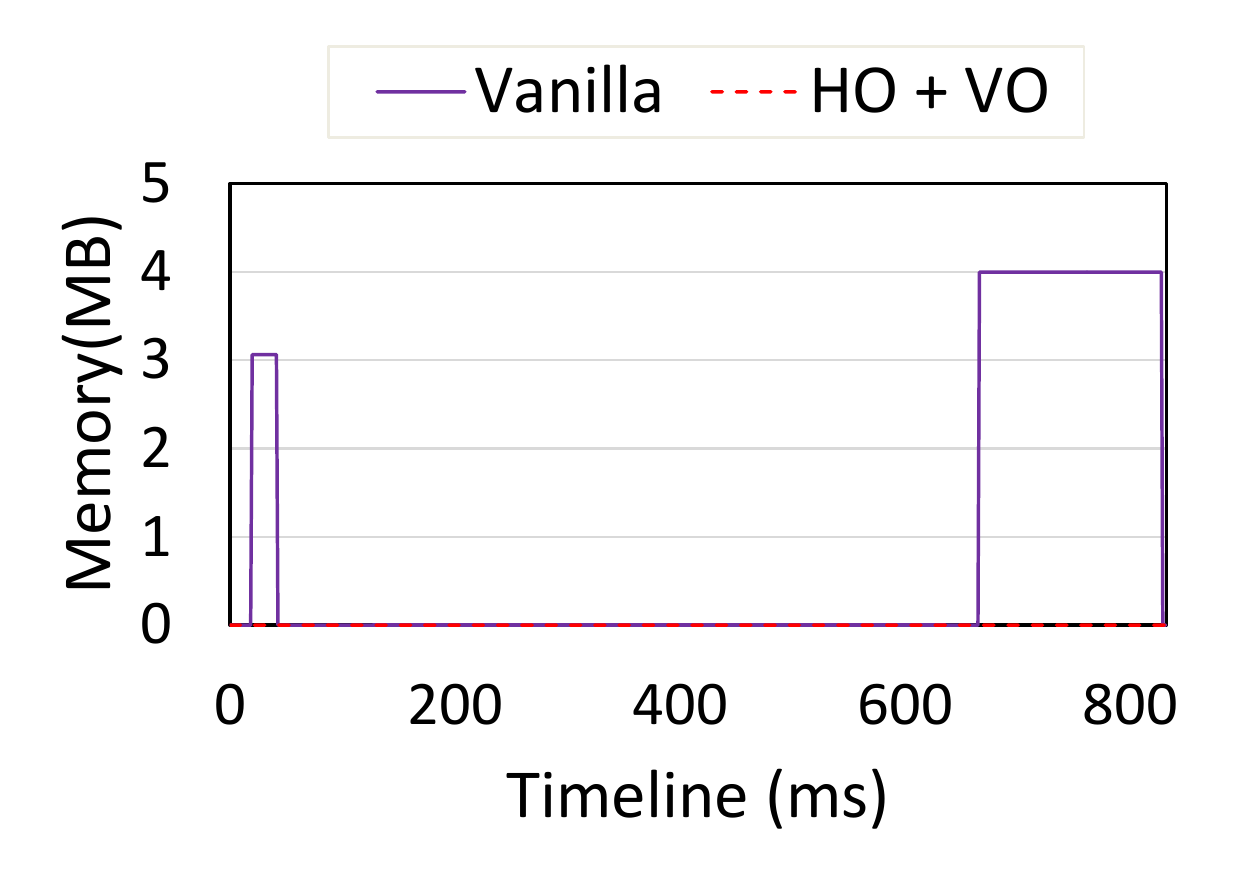}
	}
	\caption{Resources cost comparison on TMS320C6678 (MobileNet)}
	\label{MC-partial} 
	\vspace{-0.5cm}
\end{figure*}



\subsubsection{Comparison on TMS320C6678}
Figure~\ref{MC-partial} presents the resource cost comparison when running MobileNet on TMS320C6678. The other 6 models show similar trends, so we omit them due to space limitation. Since the difference in storage cost is mainly due to HO rather than VO, we omit the intermediate baseline and only compare Vanilla and the full \sysname solution.

Vanilla’s DDR overheads are contributed by both the feature maps and the parameters (e.g. weights). During the inference process, both feature map size and parameter size can go beyond the storage capacity of SRAM. As shown in Figure~\ref{Mddr}, the burst of DDR during the initial $\sim$22ms is mainly due to the output feature map. The input feature map was occupying the SRAM at that time, and SRAM becomes insufficient to hold both the input feature map and output feature map. At the end of $\sim$163ms, the burst is due to the large-sized convolution layer of MobileNet, whose parameter size reaches more than 4MB and cannot be placed in either L2 memory or SRAM-based shared memory. 

\subsubsection{Comparison on ZCU102}


\begin{figure}[!htbp]
\vspace{-0.6cm}
	\centering
	\includegraphics[width=0.49\textwidth]{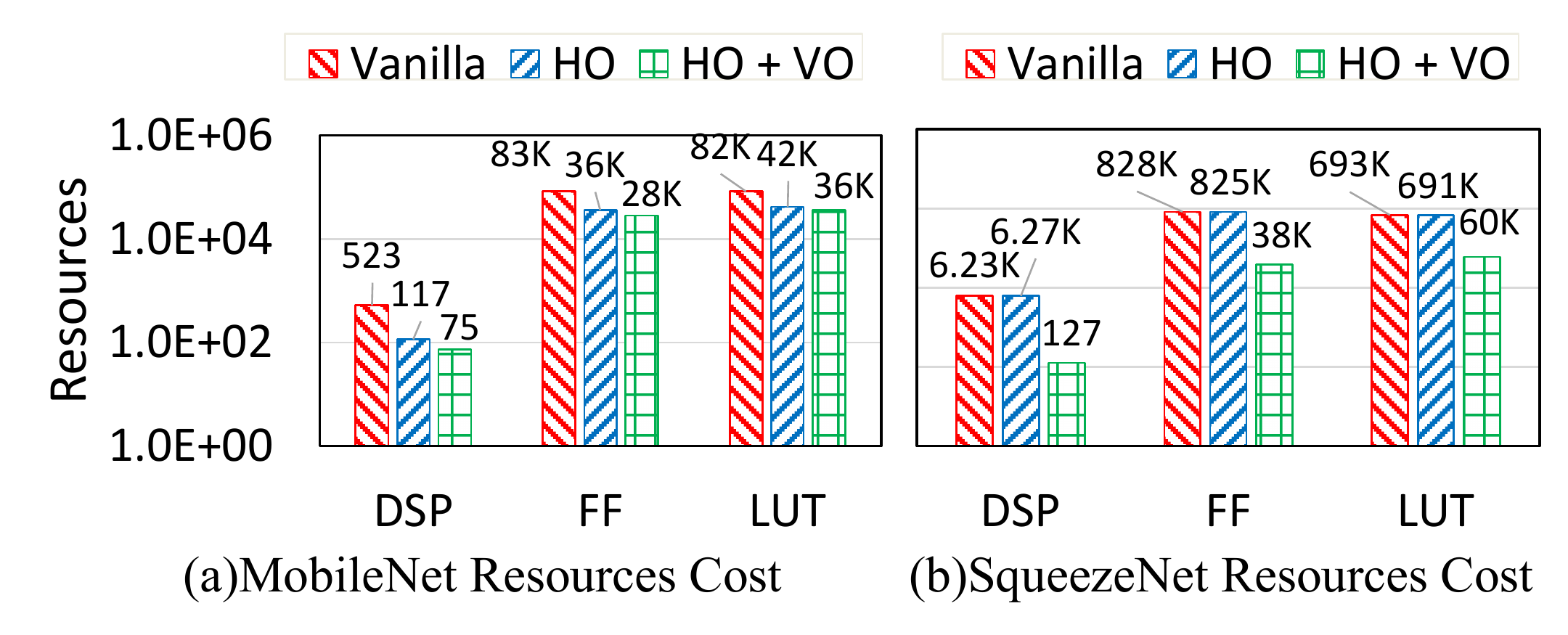}
	\vspace{-0.6cm}
	\caption{Resources cost comparison on ZCU102}
	\label{Restructure-partial}
	\vspace{-0.3cm}
\end{figure}
Figure~\ref{Restructure-partial} illustrates the resource cost of MobileNet and SqueezeNet on ZCU102 respectively, and we can see that both HO and VO help to reduce the resource cost on ZCU102. The other 5 models shows similar trend as MobileNet and thus are omitted.

HO enables higher parallelism and improves resource utilization efficiency. The computation of some inference operations are completed faster, so that the corresponding resource (e.g. DSP units) can be freed and reused for the other inference operations, instead of allocating more resource. Compared with Vanilla, HO can complete the same workload with shorter time but with less resource cost.

VO also contributes to the reduction of the resources cost. Because of the restructured dataflow, the data access becomes faster, which in turn reduces the idle time of resources (e.g. DSP units do not need to wait for the input feature maps for long time). Therefore, the resource utilization is improved and the inference engine does not need to request unnecessarily more resource from the edge device.

However, we notice that SqueezeNet implies some inconsistent trend with the other models regarding the resource cost. More specifically, HO does not help to reduce the cost of DSP units (even leads to a slight increase). We finally identify that SqueezeNet's network structure can be easily paralleled, which makes it benefit from the default optimization from HLS. In other words, HLS also integrates some optimization mechanism during the code generation, which already helps Vanilla to achieve a high utilization of DSP units, leaving little optimization room for HO. As a result, HO bring no evident performance benefit towards SqueezeNet.

\subsection{Distributed \sysname (d-\sysname)}
\label{sec-dxenos}

\begin{figure}[]
	\centering
	\includegraphics[width=0.45\textwidth]{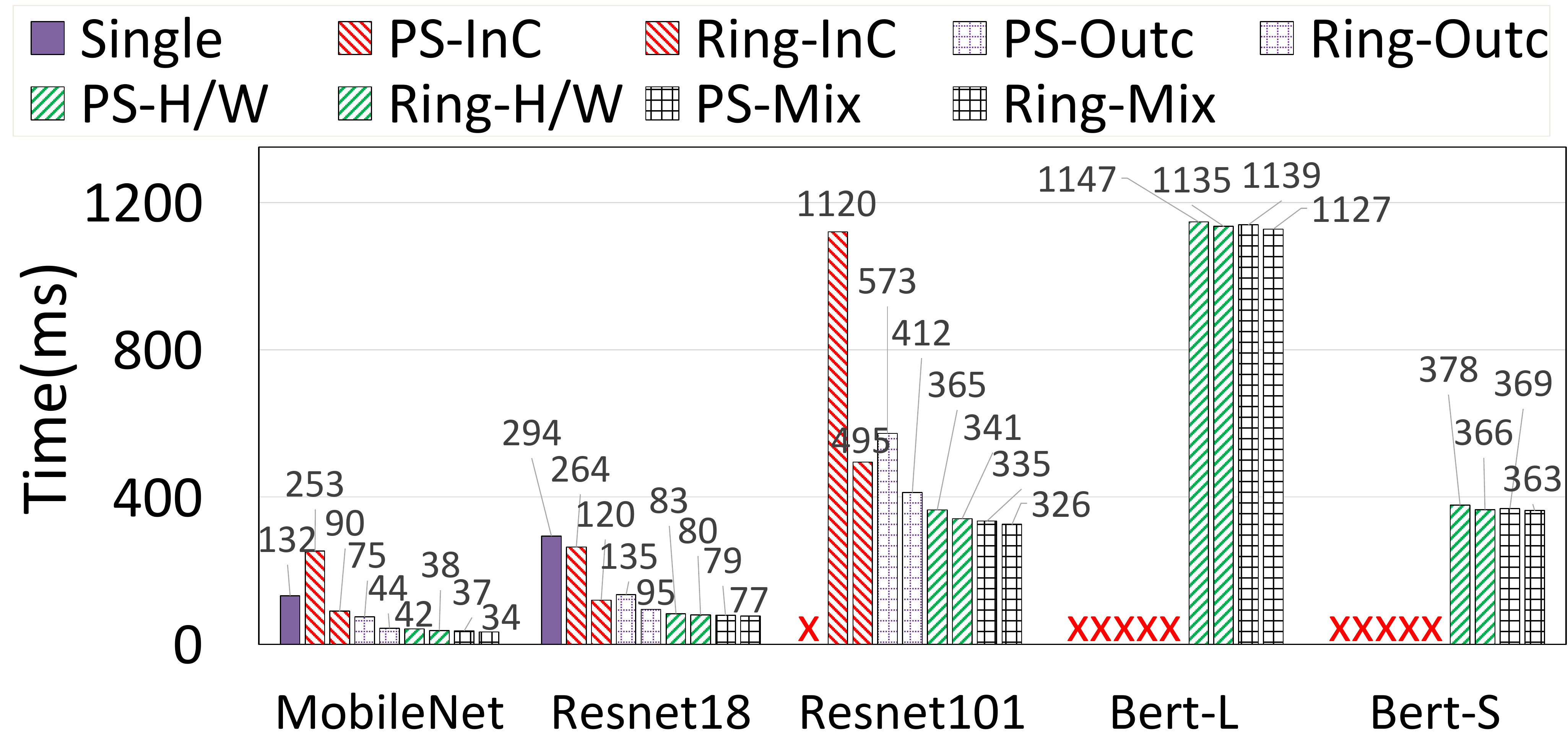}
	\caption{\texttt{d-}\sysname Inference time}
	\label{DXenos}
	\vspace{-0.6cm}
\end{figure}


 We employ 4 TMS320C6678 devices to evaluate \dsysname.  We equip \dsysname with the ring all-reduce synchronization algorithm, which proves to be bandwidth optimal in parameter synchronization~\cite{ring-allreduce}. We run three models (MobileNet ResNet, and Bert) in \dsysname. The evaluation result is shown as Figure~\ref{DXenos}~\footnote{Since Bert-L and Bert-S model only has one channel and thus do not have $inC-$/$outC-$based partition scheme. The single-node inference of ResNet-101/Bert-L/Bert-S costs much longer time than the other baselines, so we do not include them in Figure~\ref{DXenos}.}.  We summarize two main takeaways: 
 
 (1) The ring all-reduce synchronization is more efficient than the traditional PS-based synchronization. When equipped with PS-based synchronization, the inference time can become even worse compared with the single-device inference because parameter synchronization dominates the overheads. By contrasts, when equipped with ring all-reduce synchronization, \dsysname achieves distinct speedup. 
 
 (2) There is no ``one-size-fits-all'' partition scheme to achieve optimal inference time. Therefore, none of the single-mode partition schemes (i.e. \emph{inH}-based. \emph{inW}-based and \emph{outC}-based) achieves the optimal performance. By contrast,  our profiling-driven method chooses the most appropriate partition schemes for the related operators, thus making the hybrid partition scheme (Ring-Mix) yield the most performance advantage for \dsysname.



\section{Discussion: \sysname vs. TASO and PET}
\label{black-vs-white}

TASO~\cite{sosp19_taso} and PET~\cite{osdi21_pet} are considered as the most relevant works to \sysname. Knowledgeable readers may wonder, \emph{would TASO and PET be able to search out a comparable (or even better) solution than \sysname, if they were equipped with the same operator library?} Theoretically, if TASO or PET were equipped with a proper cost function, and were given infinite time and computation power, their search algorithm (i.e. essentially depth-first-search, DFS) should output a scheme, which can yield an optimal memory layout and model partition. However, such cases are impractical due to two reasons. First of all, the cost function, though claimed to be customized, is hard to define for memory layout. TASO and PET simply use the execution time as the cost function, and there are no guidelines on defining a cost function targets at memory layout, so it cannot preserve the data locality as \sysname's vertical optimizaton does. More importantly, the search algorithms of TASO and PET are simply based on the enumeration of all candidates with DFS, without considering any prior knowledge of the hardware platform. Therefore, its search space can easily blow up. Even after pruning, the search-based optimization can only work with a very small number of operators. To be more specific, TASO can only execute its search-based optimization with at most 4 operators, whereas PET can work with at most 5 operators in practice, thus constraining their application scope.

\sysname, on the other hand, leverages the prior knowledge--including the resource information of the hardware and the dataflow information of the operators--as its guideline, so its horizontal and vertical optimization avoid the explosive search space and can find a near-optimal scheme efficiently. However, TASO/PET's and \sysname's optimization approaches are not mutually exclusive and the automatic search algorithm from TASO/PET can also be inherited by \sysname to discover more optimized schemes. We leave the integration of such optimization methods as our future work.









\section{RELATED WORK}
\label{sec-related}

\para{Graph-level optimization}
Existing machine learning frameworks represent neural network models as computation graphs, and perform graph-level optimization to optimize machine learning tasks.
TensorFlow~\cite{abadi2016tensorflow} with XLA~\cite{tensorflow_xla}, TensorRT~\cite{TensorRT}, MetaFlow~\cite{MetaFlow} and DNNFusion~\cite{dnnfusion} optimize the computation graph by transformation rules designed by domain experts.
TASO~\cite{sosp19_taso} and PET~\cite{osdi21_pet} further adopts super-optimization technique for automatically graph-level optimization, which can significantly enlarge the optimization space while reducing human efforts. 
Besides, when there is a lack of hardware knowledge, simply applying the above optimization techniques cannot achieve promising performance.
Thus, \sysname{} uses architecture-aware approaches (DSP-aware operator split and operator linking) to perform more in-depth optimization.

\para{Code generation}
Halide~\cite{halide} presents a domain specific programming language for tensor programs and proposes `computation+schedule' model to decouple the optimization stages.
Inspired by Halide, TVM~\cite{osdi18-tvm} uses a similar optimization model and proposes a learning-based approach to automatically search through the hyper parameters for a given schedule to generate highly efficient code.
Ansor~\cite{ansor} and FlexTensor~\cite{flextensor} explore different schedule strategies to further enlarge the searching space for TVM.

\para{Inference on edge hardware}  AOFL parallelization\cite{zhou2019adaptive} improves edge-based inference efficiency by fusing convolutional layers and dynamically selecting the optimal parallelism according to the availability of computing resources and network conditions. \sysname, by contrast, considers more about the resource (memory and computation) conditions on the edge device (HO), and brings deeper optimization to the inter-operator dataflow (VO). 
Mema\cite{galjaard2021mema} enhances the scheduling policy to run multiple inference jobs without additional edge resources. While \sysname currently focuses on accelerating single inference job, we believe its strategies can become compatible to work in multiple-job scenario with some adaption. \cite{hadidi2019collaborative} proposes a collaborative solution, which employs both cloud resource and edge devices to jointly undertake one big inference task. It would be an interesting direction for \sysname to leverage cloud resource to accelerate the edge-based inference and we leave it as our future work.

\vspace{-0.2cm}
\section{Conclusion and Future Work}
\label{sec-conc}
We present \sysname, which incorporaes dataflow-centric optimization strategies to accelerate edge-based inference.  We conduct comprehensive experiments with 7 benchmarks on two typical platforms, Multi-Core DSP(TMS320C6678) and FPGA(ZCU102). Evaluation results demonstrate the effectiveness of both \sysname' vertical and horizontal dataflow optimization, and also prove \sysname's outperformance over TVM while executing edge-based inference under the same setting. We also develop a primary distributed version (\dsysname), which can achieve a speedup of 3.68$\times$--3.78$\times$ compared with the single device.


The future development of \sysname will mainly include two aspects: (1) We will continue to optimize the distributed version of \sysname to improve the efficiency of joint inference across multiple edge devices. (2) We will consider incorporating TASO/PET's approaches to enhance the optimization strategies of \sysname.

\clearpage

\bibliographystyle{plain}
\bibliography{ref}

\end{document}